\documentclass[a4paper]{PoS}

\usepackage{graphicx,amsfonts,amssymb,amsmath,epsfig,amsthm,fancyhdr,eucal,appendix}
\usepackage{caption, subcaption}
\usepackage{fontenc, times, mathptmx}
\usepackage[english]{babel}
\usepackage{simplewick}
\usepackage{cite}
\usepackage[utf8]{inputenc}

\newcommand{\BE}{\begin{equation}}
\newcommand{\EE}{\end{equation}}
\newcommand{\mrm}{\mathrm}
\newcommand{\dd}{\mathrm{d}}

\newcommand{\me}{\mathrm{e}}
\newcommand{\mcal}{\mathcal}
\newcommand{\mn}{\mathnormal}
\newcommand{\del}{\partial}
\newcommand{\eps}{\epsilon}

\newcommand{\mPhi}{\mn\Phi}

\title{Stochastic Renormalization Group and Gradient Flow in Scalar Field Theory}

\ShortTitle{Stochastic RG and Gradient Flow in Scalar Field Theory}

\author{\speaker{Andrea Carosso}$^{\;a}$, Anna Hasenfratz$^a$, Ethan T. Neil$^{a}$\\
		$^a$Department of Physics, University of Colorado, Boulder, CO 80309, USA\\
		E-mail: \email{andrea.carosso@colorado.edu}}

\abstract{Recently, the connections between gradient flow and renormalization group have been explored analytically and numerically. Gradient flow (when modified by a field rescaling) can be characterized as a continuous blocking transformation. In this work, we draw a connection between gradient flow and \emph{functional} renormalization group by describing how FRG can be represented by a stochastic process, and how the stochastic observables relate to gradient flow observables. The relation implies correlator scaling formulae that can be applied numerically in lattice simulations. We present preliminary results on anomalous dimensions obtained from such measurements in the context of 3-dimensional lattice $\phi^4$ theory.}

\FullConference{The 37th Annual International Symposium on Lattice Field Theory - LATTICE2019\\
		16-22 June, 2019\\
		Wuhan, China.}

\begin{document}

\section{Introduction}

In the past decade, a new tool known as ``gradient flow'' (GF) has become popular in lattice simulations as a means of setting the scale \cite{Narayanan:2006rf,Luscher:2010iy}. Because GF consists in a diffusion process on the degrees of freedom of the theory, which involves a local smoothing of the field, there have arisen attempts to characterize GF as a renormalization group transformation \cite{Yamamura:2015kva, Abe:2018zdc, Carosso:2018bmz, Carosso:2018rep, Sonoda:2019ibh, Carosso:2019qpb, Hasenfratz:2019hpg}. Since GF does not automatically involve a field rescaling, however, it is clear that it cannot in itself constitute an RG transformation (for non-compact fields, at least). By supplementing GF with such a rescaling, methods have been devised to extract quantities such as critical exponents and beta functions in lattice systems \cite{Carosso:2018bmz, Carosso:2018rep,Hasenfratz:2019hpg}. The method provides an alternative to typical methods like Monte Carlo RG or Dirac eigenmode analyses, and has been put to use in various lattice theories such as 12-flavor SU(3) gauge theory and 2-flavor QCD \cite{Carosso:2018bmz,Hasenfratz:2019hpg}.

The fact that GF is a continuous transformation suggests that there is a relation to the enterprise of functional RG \cite{Rosten:2010vm}, wherein continuous, nonperturbative RG transformations are defined, taking the form of diffusions on the space of actions. As described in \cite{Carosso:2019qpb}, it can be demonstrated that the FRG equations are equivalent to Fokker-Planck equations, and may therefore be represented by stochastic processes. The stochastic observables, in turn, may be shown to approach GF observables, at least in the limit of large separations among the operators in the expectation values. Furthermore, one may derive RG scaling relations in this formalism which parallel those of standard spin-blocking techniques, and these relations may be numerically computed in lattice simulations, as we describe below.

\section{GF and FRG}

For scalar fields $\varphi$, the gradient flow (GF) transformation is defined by a diffusion equation \cite{Luscher:2010iy}
\BE
\del_t \phi_t (x) = -\frac{\delta \hat S(\phi)}{\delta \phi(x)}(\phi_t),
\EE
where $\hat S$ may be called the ``flow action,'' and the initial condition is $\phi_0(x) = \varphi(x)$. In the simplest case of ``free'' GF, $\hat S$ is a massless, free-field action, so the GF equation is a heat equation
\BE
\del_t \phi_t (x) = \Delta \phi_t(x).
\EE
Solutions are given by integration of the initial field against the heat kernel,
\BE
\phi_t(x) = (f_t \varphi)(x) = \int_{\mathbb{R}^d} \dd^d z \frac{\me^{-z^2/4t}}{(4 \pi t)^{d/2}} \varphi(x+z),
\EE
which bears a resemblance to the usual definition of a spin blocking transformation \cite{Carosso:2018rep},
\BE
\varphi_b(n/b) = \frac{b^{\Delta_\phi}}{b^d} \sum_{\varepsilon} \varphi(n+\varepsilon),
\EE
when one identifies $b \propto \sqrt{t}$, except the scaling dimension of $\phi$ is not present in the GF solution.\footnote{This \emph{could} be implemented in the GF equation, if $\Delta_\phi$ were known ahead of time.} Since $t$ is a continuous parmeter, the GF solution is suggestive of a ``continuous'' blocking transformation. We will see below, however, that GF is more naturally identified with a different kind of RG transformation.

Continuous RG transformations have been defined in continuum field theory in the framework of functional RG (FRG) \cite{Wilson:1973jj,Rosten:2010vm}. The original approach of Wilson and Kogut was to define a low-mode Boltzmann factor by integration against a so-called ``constraint functional'' $P_t(\phi,\varphi)$ as
\BE
\me^{-S_t(\phi)} = \int [\dd \varphi] P_t(\phi,\varphi) \me^{-S_0(\varphi)},
\EE
where $S_0(\varphi)$ is the bare action of the theory of interest. $P_t$ should have the property of setting the low-mode effective field $\phi$ equal to a smoothed-out $\varphi$ within some variation, for example, via
\BE
P_t(\phi,\varphi) = N_t \exp \Bigg(- \int_p p^2 K_0^{-1}(p) \frac{[\phi(p) - f_t(p) \varphi(p)]^2}{1 - f^2_t(p)} \Bigg),
\EE
where $f_t(p)$ is the momentum space heat kernel (which damps high-modes), and $K_0(p)$ is a cutoff function, e.g. $\me^{-p^2 a_0^2}$. One can show that $P_t$ is the Green function(al) solution of the Fokker-Planck equation
\BE
\frac{\del P_t(\phi)}{\del t} = \int_p \Bigg( \frac{1}{2} K_0(p) \frac{\delta^2 P_t(\phi)}{\delta \phi(p) \delta \phi(-p)} + p^2 \phi(p) \frac{\delta P_t(\phi)}{\delta \phi(p)} \Bigg),
\EE
with initial condition $P_0(\phi) = \delta(\phi - \varphi)$. From the theory of stochastic processes, however, we may determine that $P_t$ viewed as a transition probability is generated by the Langevin equation (in particular, an Ornstein-Uhlenbeck process) on the momentum modes given by
\BE
\phi_t(p) = - p^2 \phi_t(p) + \eta_t(p),
\EE
where $\eta_t(p)$ is a regularized gaussian noise. This is similar to the free GF equation except that it is modified by a noise term.

The FRG transformation above must be supplemented with a field rescaling $b_t^{\Delta_\phi}$, which can be done passively after integrating over the bare field $\varphi$: set $\phi(p) = b_t^{\tilde \Delta_\phi} \mPhi(\bar p)$, where $\bar p = b_t p$, and $b_t$ is the continuum analog of a blocking factor, $b_t = a_t / a_0$.\footnote{$\tilde \Delta_\phi$ is the scaling dimension of the momentum space field $\varphi(p)$.} For free field theory, one can explicitly compute $b_t = \sqrt{1+2t}$ under Schwinger regularization. After this rescaling, the effective action $S_t$ may generally possess an infrared fixed point $S_*$. In the case of $\phi^4$ in 3 dimensions, it was demonstrated that indeed an IRFP exists in perturbation theory \cite{Carosso:2019qpb}. Using the definition of noise expectation values $\mathbb{E}$, one can then show that the rescaled effective observables can be written as double averages
\BE
\langle \mcal{O}(\mPhi) \rangle_{S_t} = \big\langle \mathbb{E} \big[ \mcal{O}\big(b_t^{\Delta_\phi} \phi_t(p) \big) \big] \big\rangle_{S_0}.
\EE
The right-hand side is numerically computable with Monte Carlo methods, and therefore constitutes a stochastic version of MCRG \cite{Swendsen:1979gn}, if one knows the form of the rescaling factor $b_t$. Generally, $b_t$ should satisfy $b_0 = 1, \; b_t \propto \sqrt{t}$, for sufficiently large $t$.

\section{Scaling Formulae}

The connected $n-$point functions of the effective theory may be written in terms of solutions $\phi_t$ of the GF equation using an identity found by Wilson and Kogut \cite{Wilson:1973jj}, and recently related to GF in \cite{Sonoda:2019ibh}. For 2-point functions it reads
\BE
\langle \phi(x) \phi(y) \rangle^\mrm{conn}_{S_t} = \langle \phi_t(x) \phi_t(y) \rangle^\mrm{conn}_{S_0} + A_t(x-y),
\EE
where $A_t(z)$ is a function which decays as a gaussian for $z \gg a_t$ \cite{Carosso:2019qpb}. A similar formula also holds for correlators of other local operators, such as $\phi^2(x)$. The key feature is that, at large separations, one can therefore approximate the effective correlators by GF correlators, thereby avoiding the necessity of a full Langevin simulation.

By including the field rescaling $b_t^{\Delta_\phi}$ and using the Markov property of the effective theory, one can derive scaling relations analogous to those of typical spin-blocking transformations \cite{Cardy:1996xt,Carosso:2019qpb}. Applied to correlators of arbitrary scaling operators $\mcal{O}(x)$, the scaling formula reads, for small time steps $\eps$,
\BE
\langle \mcal{O}(\bar x_1) \mcal{O}(\bar x_2) \rangle_{S_{t+\eps}} \approx b_\epsilon(t)^{2\Delta_\mcal{O}} \langle \mcal{O}(\bar y_1) \mcal{O}(\bar y_2) \rangle_{S_{t}},
\EE
where $\Delta_\mcal{O}$ is the scaling dimension of $\mcal{O}$, $b_\epsilon(t) = b_{t+\epsilon}/b_t$, and $\bar x = x / a_t$. The operators $\mcal{O}$ above are understood to be built out of rescaled fields $\mPhi(\bar x) = b_t^{\Delta_\phi} \phi(x)$, where $\phi$ is the effective field at time $t$. We then use the MCRG principle to relate the expectations on either side to the stochastic observables, and if we take large spatial separations, we arrive at the ratio formula
\BE \label{ratioformula}
R_{\mcal{O}}(t, x_2-x_1)= \frac{\langle \mcal{O}_{t+\eps}(x_1) \mcal{O}_{t+\eps}(x_2) \rangle_{S_0}}{\langle \mcal{O}_{t}(x_1) \mcal{O}_{t}(x_2) \rangle_{S_0}} \approx b_\eps(t)^{2(\Delta_\mcal{O} - m \Delta_\phi)},
\EE
where $m$ is the number of factors of $\phi$ appearing in $\mcal{O}$, and $\mcal{O}_t$ is evaluated on GF fields $\phi_t$. The scaling dimensions can be written as $\Delta_\mcal{O} = m d_\phi + \gamma_\mcal{O}$, so $\Delta_\mcal{O} - m \Delta_\phi = \gamma_\mcal{O} - m \gamma_\phi$, and the formula above may be used to find anomalous dimension differences relative to the fundamental field $\phi$.

\section{Simulation}

We simulated $\phi^4_3$ on a cube $\mathbb{Z}^3_L$ of linear sizes $L = 24, 36, 48, 56$, with the lattice action
\BE
S(\varphi) = \sum_n \Big[ - \beta \sum_{\mu = 1}^3 \varphi(n) \varphi(n+\mu) + \varphi^2(n) + \lambda (\varphi^2(n)-1)^2\Big]
\EE
using a mixed update algorithm for which one sweep consisted of one Wolff clutser update and five Metropolis updates of the radial size of $\varphi$, with maximum update length $d_\mrm{rad} = 2.00$ (see \cite{Hasenbusch:1999mw} for details about the radial update). The radial update acceptance was $\approx 0.65$. The thermalization cut was $10^4$ sweeps. Autocorrelations in the $\varphi^2$ observable were in the range $\tau_\mrm{int} \approx 4.19 - 5.34$, and errors on binned data plateau around bin size 100. Flow measurements were made every 5 sweeps. With 1 million total sweeps, we therefore have $\approx 10^4$ independent samples of flowed data. We simulated very close to criticality at $\lambda = 1.1$ using the precisely known value $\beta_\mrm{c} \approx 0.3750966$ \cite{Hasenbusch:1999mw}.

\begin{figure}[t]
\vspace{-0.5cm}
\centering
\scalebox{0.6}{
  \includegraphics[width=1.2\textwidth]{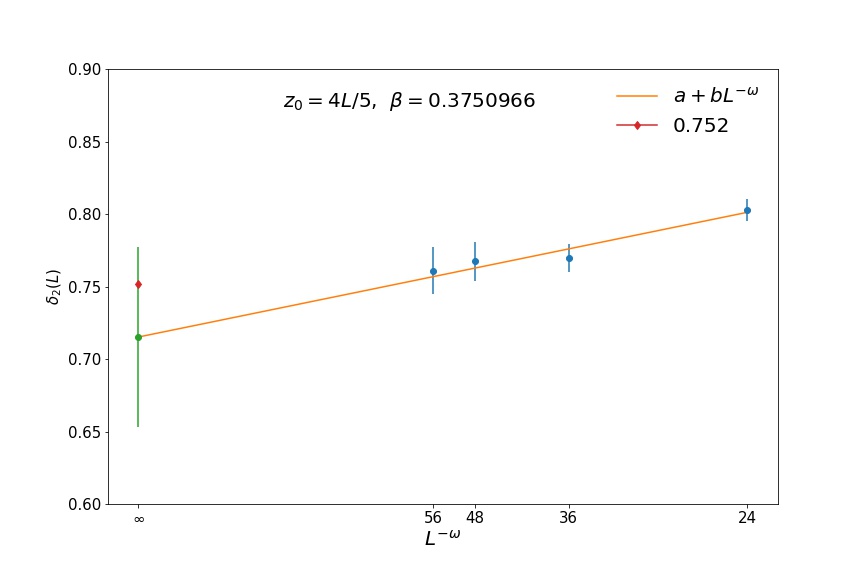}
}
\caption{\small{Infinite volume extrapolation of the difference $\delta_2(L)$ related to $\gamma_{\phi^2}$. \label{fig:delta2inf}}}
\end{figure}

By measuring correlation functions of the gradient-flowed operators $\{\phi, \phi^3\}$ and $\{\phi^2, \phi^4\}$ in the odd (even) subspaces, we tested the ratio formula, eq. (\ref{ratioformula}), under small steps $\eps = 0.01$.\footnote{Because we utilized the long-distance equivalence of gradient-flowed and effective correlators, there was no stochastic element in the simulation. A full Langevin simulation will be pursued in future work.} In the case of $\phi-\phi$ correlators, the formula actually predicts that the exponent of $b_\eps(t)$ should be 0, meaning that correlations of $\phi$ operators cannot be used to determine $\gamma_\phi$. The correlator ratio should plateau at 1 at large distances $z \gg a_t$, and this was verified (see \cite{Carosso:2018rep}). For the observable $\mcal{O} = \phi^2$, we expect to measure the difference $\delta_2 = 2( \gamma_{\phi^2} - 2 \gamma_\phi) \approx 0.752$ \cite{Hasenbusch:1999mw}. To extract $\delta_2$ from the ratios $R_\mcal{O}$, we parametrized the scale factor by $b_t = \sqrt{1 + ct}$, which implies a relative scale factor
\BE
b_\eps(t) = \Big(1 + \frac{\eps}{c^{-1} + t}\Big)^{1/2}.
\EE
On any given volume, we fit the ratios to $b_\eps(t;c)^{2\delta_2}$ over a range of $t$-values to obtain estimates for the parameters $c, \delta_2$. To obtain errors we performed a jackknife analysis, wherein the fit results $c, \delta_2$ were computed for every member of a jackknifed ensemble \cite{DeGrand:2006zz}, and the jackknife mean and variances were then obtained; these are the blue points in Fig. \ref{fig:delta2inf}. To extrapolate to infinite volume, we fit to the ansatz,
\BE \label{infvol}
\delta_2(L; a, \omega, \delta_2) = \delta_2 + b L^{-\omega},
\EE
where $\omega$ in principle should be the leading correction-to-scaling exponent, namely $\approx 0.845$. Results for these quantities are in Table \ref{fitresults}; the infinite volume estimate for $\delta_2$ was $0.715(62)$, and $\omega$ was consistent with the true value, albeit with large error.

\begin{table}[h!]
\begin{center}
 \begin{tabular}{||c c c||} 
 \hline
$\delta_2$ & $b$ & $\omega$ \\ 
 \hline\hline
 0.715(62) & 1.3(2.3) & 0.86(77) \\ 
 \hline
\end{tabular}
\caption{Fit results for the infinite volume extrapolation determined by eq. (\ref{infvol}).}
\label{fitresults}
\end{center}
\end{table}
\vspace{-0.5cm}

\begin{figure}[t]
\vspace{-0.5cm}
\centering
\scalebox{0.6}{
  \includegraphics[width=1.2\linewidth]{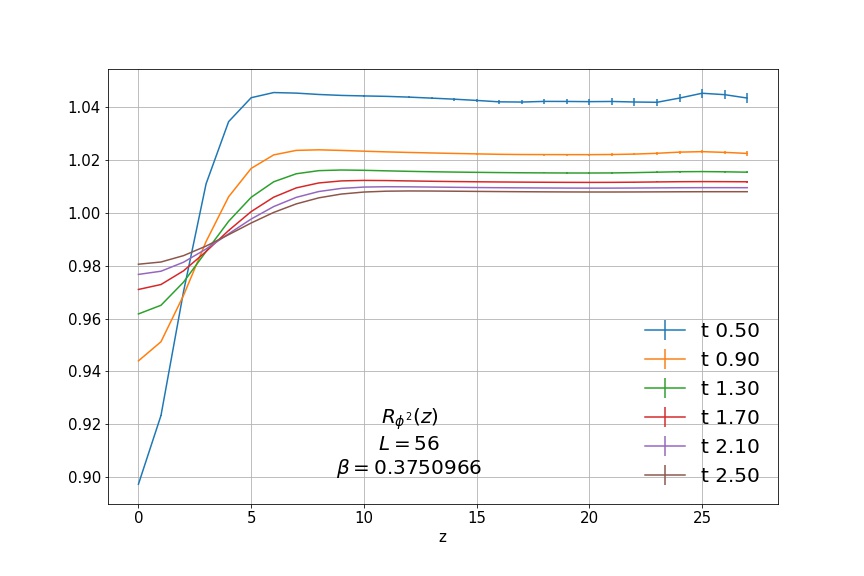}
}
\captionof{figure}{\small{$R_{\phi^2}$ as a function of separation.}}
\label{fig:phi2ratio}
\end{figure}

We note, however, that the ratio formula, strictly speaking, applies only to scaling operators, and it is known that the scaling operators are generally polynomials in $\phi$, rather than monomials. The analysis for $\phi^2$ above suggests therefore that $\phi^2$ is \emph{close} to a scaling operator, an observation justified by noting that $\phi^2$ is the leading relevant operator in the even subspace. In fact, correlators of $\phi^4$ operators displayed almost the same power law behavior as those of $\phi^2$, consistent with the fact above, suggesting that $\phi^4$ is \emph{not} a good approximation to a scaling operator. To extract dimensions of less-relevant operators, then, a more sophisticated method is required. To that end, we attempted a \emph{diagonalization} procedure in the spirit of \cite{Caselle:1999tm}. Generally, a scaling operator $\mcal{O}_a$ takes the form
\BE
\mcal{O}_a(\mPhi) = \sum_{i} c_a^i \mcal S_i(\mPhi),
\EE
where the $\mcal S^i$ are familiar (monomial) operators in the action. Conformal symmetry at a fixed point implies an orthogonality of mixed correlators \cite{Cardy:1996xt},
\BE
\langle \mcal{O}_a \mcal{O}_b \rangle = \delta_{ab}.
\EE
By choosing a basis of action operators $\mcal{B} = \{\mcal{S}_i\}$ and measuring the mixed correlators of its elements, $\langle \mcal{S}_i \mcal{S}_j \rangle$, one may numerically diagonalize the matrix thereby obtained to produce estimates for correlators of scaling operators $\langle \mcal{O}_a \mcal{O}_a \rangle$.

A preliminary attempt at doing so for the truncated basis $\mcal{B}_\mrm{o} = \{ \phi, \phi^3\}$ was performed. Since $\phi$ dominates this subspace, we found a very poor signal for the $\phi^3$ contribution. Nonetheless, once the estimate for $\langle \mcal{O}_3 \mcal{O}_3 \rangle$ was obtained, we followed the strategy outlined above to find $\delta_3 = 2(\gamma_{\phi^3} - 3 \gamma_{\phi})$. The data was insufficient to perform an infinite volume extrapolation, but better signals were found on the largest two lattices. For example,  $\delta_3(48) = 2.09(37)$ was a typical estimate on $L=48$. This value is consistent (though crude) with the prediction $\delta_3 \approx 1.928$ \cite{Rychkov:2015naa}.

\section{Discussion}

In this contribution, the connection between gradient flow and renormalization group has been outlined, and the idea of stochastic RG as a type of functional RG was described. The relation to GF observables was found to be a property of long-distance correlations of the system, and this was used to arrive at correlator scaling relations similar to those of traditional spin-blocking MCRG. It was described how one may use these relations to measure anomalous dimension differences numerically. Preliminary results for the differences $\gamma_2 - 2\gamma_1$ and $\gamma_3 - 3 \gamma_1$ were reported. The former was found with reasonable precision in the infinite volume limit, but such a limit was not obtained for the latter due to a poor signal. Moving forward, we will work to improve statistics and pin down the diagonalization method described above in order to compute the differences $\delta_3, \; \delta_4$, and we are exploring possible strategies for the computation of the fundamental anomalous dimension $\gamma_\phi$, which would require correlations of operators that do not scale anomalously.

\paragraph{Acknowledgements -} This work was supported by the U.S. Department of Energy under grant DE-SC0010005.

\bibliographystyle{JHEP}
\bibliography{RG_GF}

\providecommand{\href}[2]{#2}\begingroup\raggedright\begin{thebibliography}{10}

\bibitem{Narayanan:2006rf}
R.~Narayanan and H.~Neuberger, \emph{{Infinite N phase transitions in continuum
  Wilson loop operators}},
  \href{https://doi.org/10.1088/1126-6708/2006/03/064}{\emph{JHEP} {\bfseries
  03} (2006) 064} [\href{https://arxiv.org/abs/hep-th/0601210}{{\ttfamily
  hep-th/0601210}}].

\bibitem{Luscher:2010iy}
M.~L{\"u}scher, \emph{{Properties and uses of the Wilson flow in lattice QCD}},
  \href{https://doi.org/10.1007/JHEP08(2010)071,
  10.1007/JHEP03(2014)092}{\emph{JHEP} {\bfseries 08} (2010) 071}
  [\href{https://arxiv.org/abs/1006.4518}{{\ttfamily 1006.4518}}].

\bibitem{Yamamura:2015kva}
R.~Yamamura, \emph{{The Yang-Mills gradient flow and lattice effective
  action}}, \href{https://doi.org/10.1093/ptep/ptw097}{\emph{PTEP} {\bfseries
  2016} (2016) 073B10} [\href{https://arxiv.org/abs/1510.08208}{{\ttfamily
  1510.08208}}].

\bibitem{Abe:2018zdc}
Y.~Abe and M.~Fukuma, \emph{{Gradient flow and the renormalization group}},
  \href{https://doi.org/10.1093/ptep/pty081}{\emph{PTEP} {\bfseries 2018}
  (2018) 083B02} [\href{https://arxiv.org/abs/1805.12094}{{\ttfamily
  1805.12094}}].

\bibitem{Carosso:2018bmz}
A.~Carosso, A.~Hasenfratz and E.~T. Neil, \emph{{Nonperturbative
  Renormalization of Operators in Near-Conformal Systems Using Gradient
  Flows}}, \href{https://doi.org/10.1103/PhysRevLett.121.201601}{\emph{Phys.
  Rev. Lett.} {\bfseries 121} (2018) 201601}
  [\href{https://arxiv.org/abs/1806.01385}{{\ttfamily 1806.01385}}].

\bibitem{Carosso:2018rep}
A.~Carosso, A.~Hasenfratz and E.~T. Neil, \emph{{Renormalization group
  properties of scalar field theories using gradient flow}},
  \href{https://doi.org/10.22323/1.334.0248}{\emph{PoS} {\bfseries LATTICE2018}
  (2018) 248} [\href{https://arxiv.org/abs/1811.03182}{{\ttfamily
  1811.03182}}].

\bibitem{Sonoda:2019ibh}
H.~Sonoda and H.~Suzuki, \emph{{Derivation of a gradient flow from the exact
  renormalization group}},
  \href{https://doi.org/10.1093/ptep/ptz020}{\emph{PTEP} {\bfseries 2019}
  (2019) 033B05} [\href{https://arxiv.org/abs/1901.05169}{{\ttfamily
  1901.05169}}].

\bibitem{Carosso:2019qpb}
A.~Carosso, \emph{{Stochastic Renormalization Group and Gradient Flow}},
  \href{https://arxiv.org/abs/[1904.13057]}{{\ttfamily [1904.13057]}}.

\bibitem{Hasenfratz:2019hpg}
A.~Hasenfratz and O.~Witzel, \emph{{Continuous renormalization group $\beta$
  function from lattice simulations}},
  \href{https://arxiv.org/abs/1910.06408}{{\ttfamily 1910.06408}}.

\bibitem{Rosten:2010vm}
O.~J. Rosten, \emph{{Fundamentals of the Exact Renormalization Group}},
  \href{https://doi.org/10.1016/j.physrep.2011.12.003}{\emph{Phys. Rept.}
  {\bfseries 511} (2012) 177}
  [\href{https://arxiv.org/abs/1003.1366}{{\ttfamily 1003.1366}}].

\bibitem{Wilson:1973jj}
K.~G. Wilson and J.~B. Kogut, \emph{{The Renormalization group and the epsilon
  expansion}}, \href{https://doi.org/10.1016/0370-1573(74)90023-4}{\emph{Phys.
  Rept.} {\bfseries 12} (1974) 75}.

\bibitem{Swendsen:1979gn}
R.~H. Swendsen, \emph{{Monte Carlo Renormalization Group}},
  \href{https://doi.org/10.1103/PhysRevLett.42.859}{\emph{Phys. Rev. Lett.}
  {\bfseries 42} (1979) 859}.

\bibitem{Cardy:1996xt}
J.~L. Cardy, \emph{{Scaling and renormalization in statistical physics}}.
  Cambridge, UK: Univ. Pr. (1996) 238 p. (Cambridge lecture notes in physics:
  3), 1996.

\bibitem{Hasenbusch:1999mw}
M.~Hasenbusch, \emph{{A Monte Carlo study of leading order scaling corrections
  of $\phi^4$ theory on a three-dimensional lattice}},
  \href{https://doi.org/10.1088/0305-4470/32/26/304}{\emph{J. Phys.} {\bfseries
  A32} (1999) 4851} [\href{https://arxiv.org/abs/hep-lat/9902026}{{\ttfamily
  hep-lat/9902026}}].

\bibitem{DeGrand:2006zz}
T.~DeGrand and C.~E. Detar, \emph{{Lattice methods for quantum
  chromodynamics}}. New Jersey, USA: World Scientific (2006) 345 p, 2006.

\bibitem{Caselle:1999tm}
M.~Caselle, M.~Hasenbusch and P.~Provero, \emph{{Nonperturbative states in the
  3-D $\phi^4$ theory}},
  \href{https://doi.org/10.1016/S0550-3213(99)00333-8}{\emph{Nucl. Phys.}
  {\bfseries B556} (1999) 575}
  [\href{https://arxiv.org/abs/hep-lat/9903011}{{\ttfamily hep-lat/9903011}}].

\bibitem{Rychkov:2015naa}
S.~Rychkov and Z.~M. Tan, \emph{{The $\epsilon$-expansion from conformal field
  theory}}, \href{https://doi.org/10.1088/1751-8113/48/29/29FT01}{\emph{J.
  Phys.} {\bfseries A48} (2015) 29FT01}
  [\href{https://arxiv.org/abs/1505.00963}{{\ttfamily 1505.00963}}].

\end{thebibliography}\endgroup

\end{document}